\documentclass[12pt]{article}
\usepackage{epsfig}
\usepackage{amsthm,amsmath,amssymb,amsfonts}
\voffset= - 0.3in
\hoffset= - .6in         

\catcode`\@=11
\def\marginnote#1{}

\newcount\hour
\newcount\minute
\newtoks\amorpm
\hour=\time\divide\hour by60
\minute=\time{\multiply\hour by60 \global\advance\minute by-\hour}
\edef\standardtime{{\ifnum\hour<12 \global\amorpm={am}%
        \else\global\amorpm={pm}\advance\hour by-12 \fi
        \ifnum\hour=0 \hour=12 \fi
        \number\hour:\ifnum\minute<10 0\fi\number\minute\the\amorpm}}
\edef\militarytime{\number\hour:\ifnum\minute<10 0\fi\number\minute}

\def\draftlabel#1{{\@bsphack\if@filesw {\let\thepage\relax
   \xdef\@gtempa{\write\@auxout{\string
      \newlabel{#1}{{\@currentlabel}{\thepage}}}}}\@gtempa
   \if@nobreak \ifvmode\nobreak\fi\fi\fi\@esphack}
        \gdef\@eqnlabel{#1}}
\def\@eqnlabel{}
\def\@vacuum{}
\def\draftmarginnote#1{\marginpar{\raggedright\scriptsize\tt#1}}

\def\draft{\oddsidemargin -.5truein
        \def\@oddfoot{\sl preliminary draft \hfil
        \rm\thepage\hfil\sl\today\quad\militarytime}
        \let\@evenfoot\@oddfoot \overfullrule 3pt
        \let\label=\draftlabel
        \let\marginnote=\draftmarginnote
   \def\@eqnnum{(\theequation)\rlap{\kern\marginparsep\tt\@eqnlabel}%
\global\let\@eqnlabel\@vacuum}  }

\def\appname{Appendix}
\newcounter{app}
\def\theapp{\Alph{app}}
\def\app{\par
   \addvspace{4ex}
   \@afterindentfalse
  \secdef\@app\@dapp}
\def\@app[#1]#2{\ifnum \c@secnumdepth >\m@ne
        \refstepcounter{app}
        \addcontentsline{toc}{app}{\theapp
        \hspace{1em}#1}\else
      \addcontentsline{toc}{app}{ #1}\fi
   {\parindent \z@ \raggedright
    \Large \bf \appname~\theapp .
   \Large  \bf 
    #2}\nobreak
   \vskip 4ex   \noindent
\setcounter{equation}{0}
\def\theequation{\Alph{app}.\arabic{equation}}}
\def\@dapp#1{%
{\parindent \z@ \raggedright  \bf #1}\par\nobreak}
\def\l@app#1#2{\addpenalty{\@secpenalty}%
   \addvspace{1em plus\p@}%
   \begingroup
   \@tempdima 3em
     \parindent \z@ \rightskip \@pnumwidth
     \parfillskip -\@pnumwidth
     { \bf
     \leavevmode
     #1\hfil \hbox to\@pnumwidth{\hss #2}}\par
     \nobreak
   \endgroup}

\makeatletter
\newdimen\normalarrayskip            
\newdimen\minarrayskip               
\normalarrayskip\baselineskip
\minarrayskip\jot
\newif\ifold             \oldtrue            \def\new{\oldfalse}
\def\arraymode{\ifold\relax\else\displaystyle\fi}
\def\eqnumphantom{\phantom{(\theequation)}} 
\def\@arrayskip{\ifold\baselineskip\z@\lineskip\z@
     \else
     \baselineskip\minarrayskip\lineskip1\baselineskip\fi}

\def\@arrayclassz{\ifcase \@lastchclass \@acolampacol \or
\@ampacol \or \or \or \@addamp \or
   \@acolampacol \or \@firstampfalse \@acol \fi
\edef\@preamble{\@preamble
  \ifcase \@chnum
     \hfil$\relax\arraymode\@sharp$\hfil
     \or $\relax\arraymode\@sharp$\hfil
     \or \hfil$\relax\arraymode\@sharp$\fi}}

\def\@array[#1]#2{\setbox\@arstrutbox=\hbox{\vrule
     height\arraystretch \ht\strutbox
     depth\arraystretch \dp\strutbox
width\z@}\@mkpream{#2}\edef\@preamble{\halign \noexpand\@halignto
\bgroup \tabskip\z@ \@arstrut \@preamble \tabskip\z@ \cr}%
\let\@startpbox\@@startpbox \let\@endpbox\@@endpbox
  \if #1t\vtop \else \if#1b\vbox \else \vcenter \fi\fi
  \bgroup \let\par\relax
  \let\@sharp##\let\protect\relax
  \@arrayskip\@preamble}
\def\eqnarray{\stepcounter{equation}%
              \let\@currentlabel=\theequation
              \global\@eqnswtrue
              \global\@eqcnt\z@
              \tabskip\@centering              
              \let\\=\@eqncr
              $$%
            \halign to \displaywidth  \bgroup
             \eqnumphantom \@eqnsel
      \hskip\@centering                               
    $\displaystyle  \tabskip\z@ {##}$%
    &\global\@eqcnt\@ne \hskip 2\arraycolsep
         $ \displaystyle  \arraymode{##}$\hfil
    &\global\@eqcnt\tw@ \hskip 2\arraycolsep
         $\displaystyle\tabskip\z@{##}$\hfil
         \tabskip\@centering
    &{##}\tabskip\z@\cr}
\makeatother

\newfont{\hr}{msbm10}
\newfont{\ams}{msam10}

\font\numbers=cmss12
\font\upright=cmu10 scaled\magstep1
\def\stroke{\vrule height8pt width0.4pt depth-0.1pt}
\def\topfleck{\vrule height8pt width0.5pt depth-5.9pt}
\def\botfleck{\vrule height2pt width0.5pt depth0.1pt}
\def\Zmath{\vcenter{\hbox{\numbers\rlap{\rlap{Z}\kern 0.8pt\topfleck}\kern
2.2pt
                   \rlap Z\kern 6pt\botfleck\kern 1pt}}}
\def\Qmath{\vcenter{\hbox{\upright\rlap{\rlap{Q}\kern
                   3.8pt\stroke}\phantom{Q}}}}
\def\Nmath{\vcenter{\hbox{\upright\rlap{I}\kern 1.7pt N}}}
\def\Cmath{\vcenter{\hbox{\upright\rlap{\rlap{C}\kern
                   3.8pt\stroke}\phantom{C}}}}
\def\Rmath{\vcenter{\hbox{\upright\rlap{I}\kern 1.7pt R}}}
\def\Z{\ifmmode\Zmath\else$\Zmath$\fi}
\def\Q{\ifmmode\Qmath\else$\Qmath$\fi}
\def\N{\ifmmode\Nmath\else$\Nmath$\fi}
\def\C{\ifmmode\Cmath\else$\Cmath$\fi}
\def\R{\ifmmode\Rmath\else$\Rmath$\fi}

\def\d{\partial}

\def\bea{\begin{eqnarray}}
\def\eea{\end{eqnarray}}

\def\beq{\begin{equation}}
\def\eeq{\end{equation}}
\def\ba{\beq\new\begin{array}{c}}
\def\ea{\end{array}\eeq}
\def\be{\ba}
\def\ee{\ea}
\def\F{{\cal F}}

\def\stackreb#1#2{\mathrel{\mathop{#2}\limits_{#1}}}

\def\res{{\rm res}}

\def\N2{${\cal N}=2$}
\def\4N{${\cal N}=4$}
\def\1N{${\cal N}=1$}
\def\1N*{${\cal N}=1^*$}

\def\f{{\sf f}}

\def\beq{\begin{equation}}
\def\eeq{\end{equation}}
\def\ba{\beq\new\begin{array}{c}}
\def\ea{\end{array}\eeq}
\def\be{\ba}
\def\ee{\ea}
\newcommand{\rf}[1]{(\ref{#1})}

\begin{document}

\vspace{0.3cm}

\renewcommand{\thefootnote}{\fnsymbol{footnote}}
\begin{center}
	{\Large \bf On Krichever tau-function and Verlinde formula
		}\\
	
\end{center}

\bigskip
\begin{center}
	{\large A.~Marshakov}\\
	\bigskip
	{\em I.Krichever Center for Advanced Studies, Skoltech}\\
	{\em Department of Mathematics, HSE}\\
	{\em Theory Department of LPI}\\
	\medskip
	{\sf e-mail:\ andrei.marshakov@gmail.com}\\
\end{center}
\bigskip

\begin{center}
\begin{quotation}
\noindent
We remind the definition and main properties of the Krichever quasiclassical tau-function, and turn to the application of these formulas for recent studies of two-dimensional quantum gravity. We show, that in the case of minimal gravity it turns to be directly related with the Verlinde formula for minimal models, giving in particular case its one more direct proof. Generalizations for continuous ``complex Liouville'' theory are also briefly discussed.
\end{quotation}
\end{center}
\renewcommand{\thefootnote}{\arabic{footnote}}
\setcounter{section}{0} \setcounter{footnote}{0}
\setcounter{equation}0  \setcounter{page}0
\thispagestyle{empty}


\section{Introduction}

Progress in two-dimensional (2d) quantum gravity is one of the most important recent successes in theoretical physics. Starting with famous paper \cite{Pol81}, where it has been formulated in terms of 2d Liouville conformal theory, many important issues about this theory, crucially different from gravity in higher dimensions, have been understood due to universality and renormalizability of gravity in two dimensions. In late 80-s of last century both ``world-sheet'' \cite{KPZ,DDK} and ``target-space'' or matrix model \cite{mamo,ds} approaches were essentially formulated, moreover the last one turned a possibility to identify the correlators in many 2d gravity models with certain well-known objects from the theory of integrable systems \cite{Douglas,FKN}. 

A special role in these constructions belongs to the quasiclassical tau-function, introduced by Igor Krichever in \cite{KriW}. This tau-function played a role of generating function for the correlators in models of topological 2d gravity (and topological strings) in quasiclassical approximation, i.e. on world-sheets of spherical topology. It has been defined in pure geometric terms, using (target-space!) complex curves with some extra data, and soon after  \cite{GKMMM} appeared in the context of effective supersymmetric gauge theories in higher dimensions \cite{SW}.

In any well-defined quantum theory the higher loop corrections should be fixed by classical, or tree-level approximation. In the context of string theory or 2d quantum gravity this comes from coupled to Liouville 2d conformal field theory \cite{BPZ} on higher-genus world sheets, and is believed to be governed by so called ``topological recursion'' (see e.g. \cite{TR}). Below we do not go into these issues, since even the most essential achievements in the world-sheet approach like exact 3-point functions \cite{DOZZ,Tesch,Zam3pt} and the higher equations of motion \cite{Zam_ho} in Liouville theory allow to extract information basically only about the correlation functions on sphere and torus \cite{Sphe_corr} with ``minimal amount'' of operator insertions.
Moreover, the axioms of topological recursion are hardly applicable to the main physical (i.e. classical) contribution to the partition function, even though (implicitly or explicitly) contain some basic ingredients of the construction from \cite{KriW}. Nevertheless, the application of this procedure is now widely believed to reproduce many results for the 2d gravity theories on higher genus world-sheets, see e.g. \cite{recent}.

In this note we concentrate on the application of the Krichever construction from \cite{KriW} to the minimal 2d gravity models, and find below, that one of its main ingredients -- the residue formula for third derivatives -- in this case is closely related with the Verlinde formula \cite{Ver} from 2d conformal field theory. We believe that this observation apart of providing some elementary proofs can be extended beyond these minimal theories, and even turn to be a base for some axiomatic formulation of 2d quantum gravity, which is hardly reached directly from conformal theory on world sheets.

\section{Krichever tau-function}


The geometric origin for the Krichever tau-function is given by the Riemann bilinear identities. For example, on genus-\(g\) Riemann surface \(\Sigma _g\) with $\dim H_1(\Sigma _g)=2g$, and intersection form $A_\alpha\circ B_\beta = \delta _{\alpha\beta}$, \(\alpha,\beta=1,\ldots,g\),
there are \(g\) holomorphic differentials $\bar\d (d\omega_\alpha)=0$,  which can be normalized to the $A$-cycles
\be\label{normA}
\oint _{A_\beta}d\omega _\alpha = \delta _{\alpha\beta}
\ee
Then the period matrix
\be\label{pemat}
\oint _{B_\alpha}d\omega _\beta =  T_{\alpha\beta}
\ee
is symmetric due to
\be
\label{sypema}
0=\int_{\Sigma} d\omega_\beta\wedge d\omega_\gamma= \int_{\d\Sigma} \omega_\beta d\omega_\gamma=
\\
=\sum_\alpha\left( \oint_{A_\alpha}d\omega_\beta\oint_{B_\alpha}
d\omega_\gamma-\oint_{A_\alpha}
d\omega_\gamma\oint_{B_\alpha}d\omega_\beta \right)
= T_{\beta\gamma} - T_{\gamma\beta}
\ee
and the proof comes just from the Stokes theorem on cut Riemann surface. 

This relation allows to define a prepotential (a particular case of Krichever's tau-function) by
\be
\label{prep}
{\bf a} = \frac1{2\pi i}\oint_{\bf A} dS, \ \ \ \ \ 
{\bf a}_D = \oint_{\bf B} dS := {\d\F\over\d {\bf a}} 
\ee
on a $g$-parametric family
of Riemann surfaces $\Sigma_g$ with pair of meromorphic differentials ($dx$ and
$dy$), with the fixed periods, or a generating  
differential \(dS\) and connection  \(\nabla_{\rm mod}\) on moduli space
\be
\label{dS}
dS \propto ydx, \ \ \ \ \nabla_{\rm mod} dS = \left(\nabla_{\rm mod} y\right) dx = {\rm holomorphic}
\ee
where $y(P)=\int^P dy$, $P\in\Sigma$. 
 Integrability of \eqref{prep} follow from \eqref{sypema}, since
\be
\label{period}
{\d a^D_\alpha \over \d a_\beta} = T_{\alpha\beta} = T_{\beta\alpha} = {\d a^D_\beta \over \d a_\alpha}
\ee
and for the period matrix one gets
\be
\label{F2period}
T_{\alpha\beta} = {\d^2\F\over \d a_\alpha\d a_\beta} 
\ee
To complete the definition we need in what follows the time-variables \cite{KriW,LGGKM,TakTak}, associated with the
second-kind Abelian differentials with singularities at a point $P_0$
\be
\label{tP}
t_k = {1\over k}\res_{P_0} \xi^{-k}dS,\ \ \ k>0,
\ \ \ \ \
{\d\F\over \d t_k} := \res_{P_0} \xi^{k}dS,\ \ \ k>0
\ee
where $\xi$ is an inverse local co-ordinate at $P_0$:
$\xi(P_0)=\infty$. The consistency condition for \rf{tP} is ensured by
\be
\label{sysi}
{\d^2\F\over \d t_n\d t_k} = \res_{P_0} (\xi^k d\Omega_n)
\ee
where
\be
\label{2kA}
d\Omega_k\ \stackreb{P\to P_0}{\sim}\
d\xi^k+\dots,\ \ \ \ \oint_{\bf A} d\Omega_k = 0,
\ \ \ \ \ k\geq 1
\ee
is the second kind Abelian differential. The r.h.s. of \eqref{sysi} is
symmetric due to $(\Omega_n)_+= \xi^n$, for the main, singular at $P_0$, part. Also 
\be
\label{Fta}
{\d^2\F\over \d t_n\d a_\alpha} = \oint_{B_\alpha} d\Omega_n = \res_{P_0} \xi^{n}d\omega_\alpha
\ee
which again follows from the Riemann bilinear relations, now including also the second kind Abelian differentials.


The third derivatives of the Krichever tau-function are universally given by the formula \cite{KriW}
\be
\label{Res}
{\d^3 \F\over \d \mathtt{T}_I\d \mathtt{T}_J\d \mathtt{T}_K} =
\res_{dx=0}\left(d\mathtt{H}_Id\mathtt{H}_Jd\mathtt{H}_K\over dx dy\right)
\ee
for the flat co-ordinates \(\{\mathtt{T}_I\}=\{a_\alpha,t_n,\ldots\}\) and corresponding differentials \(\{d\mathtt{H}_I\}=\{d\omega_\alpha,d\Omega_n,\ldots\}\). The proof is actually based on the property of the connection \eqref{dS}. If \(x\) is chosen to be covariantly constant, derivatives of \eqref{F2period}, \eqref{sysi} and \eqref{Fta} are localized at the points where \(dx=0\).

\section{dKP for minimal gravity
\label{ss:dkppq}}

According to common hypothesis, the $(p,q)$ critical points of
2d quantum gravity (or $(p,q)$ minimal string theory) are most effectively described by
the tau-function of $p$-reduced KP hierarchy, satisfying string equation \cite{Douglas,FKN}. The logarithm
of this tau-function should be further expanded around
certain background values \(\{t_k\}\) of the time-variables (see below), with necessary $t_{p+q}\neq 0$.
In particular, it means that the correlators on world-sheets of
spherical topology are governed by quasiclassical
tau-function of dispersionless KP (also called dKP) hierarchy,
which is a very particular case of generic quasiclassical tau-function from \cite{KriW}.

The geometric formulation of results for minimal string theories in terms of
the quasiclassical hierarchy can be sketched in the following way:

\begin{itemize}
	\item  For each $(p,q)$-th point (with co prime \(p\) and \(q\)) take a pair of polynomials
	\be
	\label{polspq}
	x=\lambda^p+\dots,\ \ \ \ \ 
	y=\lambda^q+\dots
	\ee
	of powers $p$ and $q$ respectively. They can be thought of as a dispersionless version of the Lax
	and Orlov-Shulman operators \cite{Douglas} of KP theory
	\be
	\left[{\hat x},{\hat y}\right]=\hbar
	\\
	{\hat x} = \d^p + \ldots + X_0,
	\ \ \ \
	{\hat y} = \d^q + \ldots + Y_0
	\ee
	or as a pair of (here already integrated) Krichever differentials \(dx\) and \(dy\) with the fixed (here vanishing)
	periods on a complex curve (for dKP - a rational curve with global uniformizing
	parameter $\lambda$). 
	\item 	The variables of dispersionless KP hierarchy are introduced by formulas \eqref{tP}
	where
	\be
	\label{dSpq}
	dS = ydx = p\lambda^{p+q-1}d\lambda + \ldots
	\ee
	and
	\be
	\label{locop}
	\xi = x^{1\over p} = \lambda\left(1+
	\dots + {X_0\over\lambda^p}\right)^{1\over p}
	\ee
	is the distinguished inverse local co-ordinate at the
	point $P_0$, where $\lambda(P_0)=\infty$ and $\xi (P_0) = \infty$.
	\item An invariant way to look at the basic polynomials \rf{polspq} is to say, that they satisfy an algebraic equation
	\be
	\label{pqcurve}
	y^p - x^q - \sum f_{ij}x^iy^j = 0
	\ee
	with some particular coefficients $\{ f_{ij}\}$. Generally, for arbitrary coefficients
	this is a smooth curve of genus~\footnote{Interpretation of this curve in terms of open strings and branes can be found in \cite{SeShi}.}
	\be
	g = {(p-1)(q-1)\over 2}
	\ee
	which is a resolution or desingularization of the situation, when $x$ and $y$ are polynomials of a global variable $\lambda$, see Fig.~\ref{fi:cu23}. This number coincides
	with the number of primaries in corresponding minimal conformal $(p,q)$ theory \cite{BPZ}. 
	\end{itemize}

The residue formula \eqref{Res} we need in what follows is therefore reduced to
\be
\label{residue}
{\d^3 \F\over \d t_k\d t_l\d t_n} =
\res_{dx=0}\frac{dH_kdH_ldH_n}{ dx dy}
\ee
The set of one-forms here
\be
\label{smer}
dH_k={\d dS\over \d t_k} , \ \ \ k\geq 1
\ee
(derivatives are taken at fixed $X$)
corresponds to dispersionless limit of KP flows and can be integrated up to
polynomial expressions
\be
\label{Hpol}
H_k = x(\lambda)^{k/p}_+
\ee
in uniformizing co-ordinate $\lambda = H_1$. Note also, that:

\begin{itemize}
	\item The formulation of minimal \((p,q)\)-models, playing the role of matter in minimal gravity, is totally symmetric under exchange of \(p\leftrightarrow q\).
	\item When coupled to 2d gravity this symmetry becomes obviously broken, already at the level of choosing \(p\)- or \(q\)- reduction of the KP hierarchy.
	\item We always consider the case of \(p\)-reduction, where independence over \(p\mathbb{Z}_{>0}\)-times is obvious, while independence over  \(q\mathbb{Z}_{>0}\)-times is realized ``implicitly''. For comparing with the world-sheet approach we imply below that \(p>q\).
	\item The generating functions of \(pq\)-dual theories are related by nontrivial transformation \cite{FKNpq,KhMa}. In terms of the Krichever data this is a kind of Fourier or Legendre transform \(dx\leftrightarrow dy\), in particular the dual to \eqref{residue} 3-point function \(\F^D_{ijk}\) is localized at the zeroes \(dy=0\). For example, 
	\be
	\label{residueD}
	{\d^3 \F^D\over \d t^D_k\d t^D_l\d t^D_n} =
	\res_{dy=0}\frac{dH^D_kdH^D_ldH^D_n}{ dx dy}
	\ee
	where \(t_k=t^D_k +\ldots \) up to polynomial resonance corrections. In the extreme case of ``empty'' \((p,1)\)-models all \(\F^D_{ijk}=0\).
\end{itemize}

\section{Conformal background}

\subsection{KPZ and conformal dimensions}

In minimal gravity one deals with the  \((p,q)\) minimal model of 2d CFT \cite{BPZ} with the central charge \(c=1-6\frac{(p-q)^2}{pq}\) interacting with the 2d Liouville gravity with the central charge  \(c_L=26-c=1+6\frac{(p+q)^2}{pq}\). 

Without loss of generality we always consider the case of \(p\)-reduction and fix \(p>q\)~\footnote{Actually the set of times \(\Pi_{p,q}\) itself is invariant under \(pq\)-duality, and we make this choice just for convenience.}. The times or couplings can be parameterized by \(\{k_{r,s} \}\in\Pi_{p,q}\) with
\be
k_{r,s} = |rq-sp|= rq-sp, \ \ \ r=1,\ldots,p-1,\ \ \ s=1,\ldots,q-1,\ \ \ \ rq-sp>0
\ee 
The sense of such numeration is rather clear, since it just corresponds to the whole set of KP times \(\{T_n|n\in \mathbb{Z}_{>0}/p\mathbb{Z}_{>0}/q\mathbb{Z}_{>0}\}\) modulo \(p\)- and \(q\)- reductions, together with their ``descendants'' (in \(p\)-reduced theory), whose number is exactly \(|\Pi_{p,q}|=\frac12(p-1)(q-1)\), being the number of primaries in the spectrum of  \((p,q)\) minimal model with conformal dimensions
\be
\label{hCFT}
h_{r,s} = \frac{k_{r,s}^2-(p-q)^2}{4pq} =  \frac{(rq-sp)^2-(p-q)^2}{4pq}
\ee 
The KP times with \(1\leq k\leq p-q\) correspond to the CFT primaries with non-positive \(h\leq 0\) dimensions \eqref{hCFT}, for \(k>p-q\) for the CFT dimensions one has \(h>0\) and for \(k\geq p+q\) we find \(h\geq 1\).

The ``background'' values for these times are generally
\be
T_n = t_n,\ \ \ 1\leq n\leq p+q-1,\ \ \ \ T_n=0,\ \ \ \ n>p+q
\\
t_{p+q} = \frac p{p+q} 
\ee 
The ``maximal'' KP time is \(T_{\rm max} = T_{pq-p-q}\), which corresponds to \(k_{\rm max} =k_{p-1,1}\).
In the ``conformal background'' one puts
\be
t_k = \frac p{p+q}\delta_{k,p+q} + \mu \delta_{k,p-q} 
\ee 
where \(\mu\) is the world-sheet cosmological constant.

Under the scaling $X\to\Lambda^p X$, $Y\to\Lambda^q Y$,
(induced by $\lambda\to\Lambda\lambda$ and 
$\xi\to\Lambda\xi$), the times \rf{tP}
transform as $t_k \to\Lambda^{p+q-k}t_k$, which corresponds (up to a numeric factor) to the KPZ dimensions of the operators from 2d CFT approach  \cite{KPZ}. 
From the second formula
of \rf{tP} it follows that the function $\F$ scales as $\F\to\Lambda^{2(p+q)}\F$,
or, for example, as
\be
\label{scaF}
\F \propto t_1^{\ 2{p+q\over p+q-1}}\ f(\mathtt{t}_k)
\ee
where $f$ is supposed to be a scale-invariant function of corresponding
dimensionless ratios of the times $\mathtt{t}_k=t_k/t_1^{p+q-k\over p+q-1}$.
Another natural scaling is of the form
\be
\label{scakdv}
\F \propto \mu^{1+\frac{p}q}\ \f({\sf t}_l)
\ee
with ${\sf t}_l = t_{l}/\mu^{(p+q-k)/2q}$, where the role of cosmological
constant is played by \(\mu = t_{p-q}\propto \Lambda^{2q}\) (for \(p>q\)).
For the \((p,q)=(2K +1,2)\) series this can be illustrated in a table 
\begin{center}
	\begin{tabular}{|c||c|c|c|c|c|c|}
		\hline
		dKP\ \(\{t_k=t_{2l+1}\}\) & \(t_1\) & \(t_3\) & \(\ldots\) & \(t_{2K -1}\) & \(t_{2K +1}\) & \(t_{2K +3}\) \\
		\hline
		CFT	Tachyons  &  \(\tau_{K -1}\) & \(\tau_{K -2}\) & \(\ldots\) & \(\tau_0\) & \(\tau_{-1}\) &  \(\tau_{-2}\)
		\\
		\hline 
KPZ dim \(=\frac{p+q-k}{2}\) &  \(K +1\) & \(K \) & \(\ldots\) & \( 2 \) & \(1\) & \( 0 \)
		\\
		\hline
		\(h_{K-l,1}\equiv h(2l+1)=-\frac{(K+l)(K -l-1)}{2(2K +1)}\)	&  \(-\frac{K (K -1)}{2(2K +1)}\) & \(-\frac{(K +1)(K -2)}{2(2K +1)}\) & \(\ldots\) & \(0\)& \(\frac{K}{2K+1}\) & \(1\)
		\\
		\hline 
	``reflected''	CFT Tachyons	& \(\tau_{K }\) & \(\tau_{K +1}\)& \(\ldots\) & \(\tau_{2K -2}\) & \(\)  &
		\\
		\hline 	
	\end{tabular}
\end{center}

\subsection{Chebyshev curves}

\begin{figure}[tp]
		\begin{center}
		\includegraphics[width=7cm]{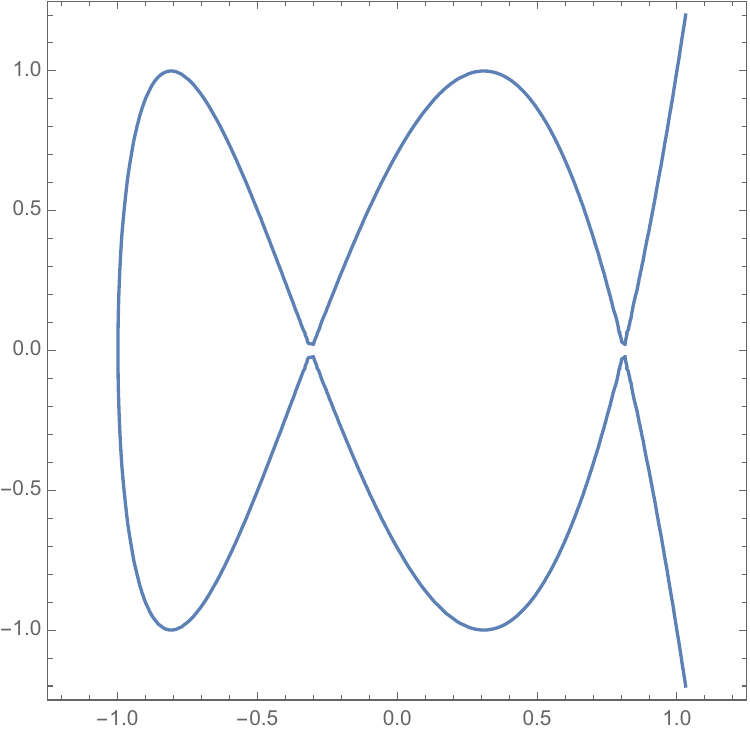}
		\hspace{0.5cm}
		\includegraphics[width=7cm]{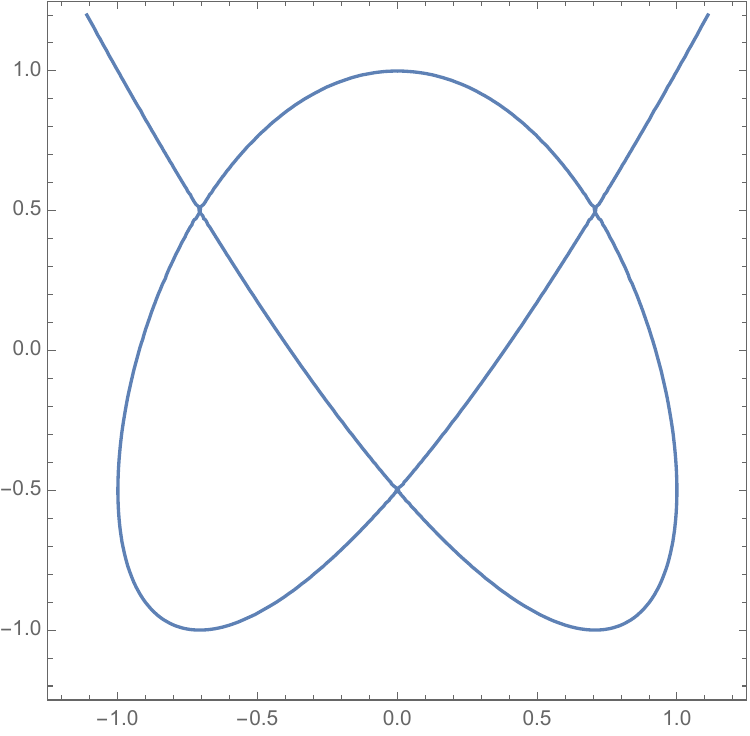}
		\caption{Degenerate curves of  Yang-Lee and Ising models of  $g=2$ and $g=3$.}
		\label{fi:cu23}
	\end{center}
	\end{figure}

At only \(\mu=t_{p-q}\neq 0\) (in addition to \(t_{p+q}=\frac{p}{p+q}\)) the spectral curve equation \rf{pqcurve} can be re-written as
\be 
\label{Chebcu}
T_p(Y) = T_q(X)
\ee 
in terms of the Chebyshev polynomials of the first type, (see Fig.~\ref{fi:cu23} for the examples of $(p,q)=(5,2)$ and $(p,q)=(4,3)$), and equivalently it means for \rf{polspq}, that
\be 
\label{Chebuni}
X=T_p(z),\ \ \ \ 
Y = T_q(z)
\ee 
with 
\be
z=\lambda/2,\ \ \ X=x/2,\ \ \ Y=y/2
\ee
compare to \eqref{polspq}. The curve \rf{Chebcu} is degenerate at the points, where
\be 
\label{degXYpq}
T_p'(Y) = pU_{p-1}(Y)= 0,
\ \ \ \ \ 
T_q'(X) = qU_{q-1 }(X)=0
\ee
which together with \rf{Chebcu} give a relation 
\be 
\label{ringXpq}
U_{p-2}(Y) =U_{q-2}(X)
\ee
from the ground ring \cite{SeShi}.

For the curves \eqref{Chebuni} the p-reduced KP Hamiltonians acquire the form
\be
H_k(z) = \left(2T_p(z)\right)^{k/p}_+  + \ldots = 2 T_k(z) 
\ee 
where the corrections disappear for \(k<2p\). We actually conjecture, that this formula is valid in conformal background for all \(\{k_{r,s} \}\in\Pi_{p,q}\), i.e. that KP times and Hamiltonians are subjected to certain triangular change of variables, which does not influence the conformal background as well as generic ``small phase space'' of \cite{KriW}. Then one finds that 
\be
\label{fimm}
\phi_n(z) = \frac1{2n}\frac{dH_n(z)}{dz} = U_{n-1}(z)\ \stackreb{z=\cos\theta}{=}\ \frac{\sin n\theta}{\sin\theta}
\ee 
and we use in what follows these formulas for any \(n\).

In the case \((p,q)=(2K +1,2)\) equations \eqref{degXYpq} and \eqref{ringXpq} read
\be 
\label{degXY1}
\frac{1}{2}T_2'(X) = U_1(X)= X=0,
\ \ \ \
\frac{1}{2K +1}T_{2K +1}'(Y) = U_{2K }(Y)=0
\\
U_{2K -1}(Y)=1
\ee
and, using uniformization \rf{Chebuni}, equations \rf{degXY1} can be easily solved, giving \(K \) nodal points on the curve with
\be 
\label{deg_sol}
\zeta^\pm_n = \pm \cos\frac{\pi (2n-1)}{2(2K +1)},\ \ \ \ n=1,2,\ldots,K 
\\
Y_n = T_2(\zeta^\pm_n) = \cos\frac{\pi (2n-1)}{2K +1},\ \ \ X_n=T_{2K +1}(\zeta^\pm_n) = \pm \cos\pi \left(n-\frac12\right) = 0
\ee 
The last relation from \rf{degXY1} is satisfied in the points \rf{deg_sol} due to
\be
U_{2K -1}(Y_n) = \frac{\sin \frac{2\pi K  (2n-1)}{2K +1}}{\sin\frac{\pi  (2n-1)}{2K +1}}
= \frac{\sin \left(\pi (2n-1) - \frac{\pi  (2n-1)}{2K +1}\right)}{\sin\frac{\pi  (2n-1)}{2K +1}} =
\\
= - \cos \pi (2n-1) = 1,\ \ \ \ n=1,2,\ldots,K 
\ee 
This is a particular case of more general ``reflection property''
\be 
\label{reflU}
U_{k-1}(Y_n) = U_{2K -k}(Y_n),\ \ \ \ k,n=1,2,\ldots,K 
\ee 
satisfied due to
\be 
\sin \frac{\pi k (2n-1)}{2K +1} = \sin \left(\pi (2n-1) - \frac{\pi  k(2n-1)}{2K +1}\right) = \sin \frac{\pi (2K +1-k) (2n-1)}{2K +1}
\ee 
In addition to \(K \) nodal points \eqref{deg_sol} on the curve \eqref{Chebcu} there is a single branch point \(z=0\) where \(dY=0\) and \(2K \) points with \(dX=0\) with
\be 
\label{dYcrit}
U_{2K }(z) = 0,\ \ \ \ z=\mathtt{z}^\pm_m = \pm \cos\frac{\pi m}{2K +1},\ \ \ \ m=1,2,\ldots,K 
\\
Y_m=T_2(\mathtt{z}^\pm_m) = \cos\frac{2\pi m}{2K +1},\ \ \ \ \ X^\pm_m=\pm T_{2K +1}(\mathtt{z}^\pm_m)=\pm\cos \pi m = \mp (-1)^m
\ee

\subsection{Examples}

\subsubsection{Tachyons and ground ring in \((2K+1,2)\) series}

Here the polynomials \eqref{fimm} correspond to the tachyonic operators
\be
\mathcal{T}_k \simeq  \frac{dH_{2K +1-2k}}{dz} = U_{2(K -k)}(z),\ \ \ \ k=1,\ldots,K 
\ee 
which, according to \cite{SeShi} form a non-faithful module of the ground ring:
\be
\label{ringU} 
U_k(Y)U_l(Y) = \sum_{m=|k-l|:2}^{k+l} U_m(Y) = U_{k+l}(Y) + U_{k+l-2}(Y) + \ldots + U_{|k-l|}(Y),\ \ \  k,l=0,1,\ldots
\ee 
modulo \(U_{2K }(Y)=0\),
which implies, in particular, ``reflection relations''
\be 
\label{refl1}
U_{2K +k}(Y) + U_{2K -k}(Y)=0
\ee 
It can be represented as algebra of functions on \(2K \) points \eqref{dYcrit}. In the tachyonic  module the ground ring operators \eqref{ringU}
act by multiplication on \(U_{k-1}(Y)=U_{k-1}(T_2(z))\), i.e.
\be
\mathcal{T}_k(z) = U_{k-1}(X)\mathcal{T}_1(z),\ \ \ \ \ k=1,\ldots,2K 
\ee 
and one can check the consistency directly by
\be 
U_{k-1}(T_2(z))U_{2K -2}(z)=\frac1z U_{2k-1}(z)U_{2K -2}(z)\ \stackreb{\eqref{ringU}}{=}\ 
\\
= \frac1z\left(U_{2K +2k-3}(z)+U_{2K +2k-1}(z)+\ldots + U_{2K -2k+3}(z) + U_{2K -2k+1}(z) + U_{2K -2k-1}(z)\right)
\ \stackreb{\eqref{refl1}}{=}\ 
\\
= \frac1z\left(U_{2K -2k+1}(z) + U_{2K -2k-1}(z)\right)\ \stackreb{\eqref{ringU}}{=}\ U_{2(K -k)}(z)
\ee 
Also, since
\be
U_{2K -k}(Y) = U_{2K -k}(T_2(z)) = \frac1z  U_{4K -2k-1}(z)\ \stackreb{\eqref{refl1}}{=}\ - \frac1z U_{2k-1}(z) = 
\\
= - U_{k-1}(T_2(z)) = - U_{k-1}(Y)
\ee 
the tachyonic operators \(\mathcal{T}_k \sim \mathcal{T}_{2K -k} \) are identified up to a normalization constant.

\subsubsection{Ising model (p,q)=(4,3)}


The residue formulas for the polynomials
\be
\label{ispol}
x = \lambda^4 + X_2\lambda^2 + X_1\lambda + X_0,\ \ \ \ y = \lambda^3 + Y_1\lambda + Y_0
\ee
give rise (for \(t_3=t_4=t_6=0\)) to
\be
\label{coef34}
Y_0= {3\over 4}X_1,\ \ \ \ Y_1 = \frac14 (5 t_5 + 3 X_2)
\\
X_0 =\frac1{24} (3 X_2^2-10 t_5 X_2)
\ee
while $X_1$ and $X_2$ satisfy
\be
\label{seis}
t_1=\frac1{48} (18 X_1^2 + 25 t_5^2 X_2 - 3 X_2^3)
\\
t_2=\frac18 (3 X_1 X_2-5 t_5 X_1)
\ee
At \(t_2=t_5=0\) under \(X_1=0, X_2=-4\) these formulas turn into the Chebyshev T-polynomials
\be
\label{Pafn_Ising}
Y=\left.y/2\right|_{\lambda=2z} =  T_3(z),\ \ \ \ X=\left.x/2\right|_{\lambda=2z}= T_4(z) 
\\
T_3(X)=T_4(Y)
\ee 
The ground ring of gravitationally dressed Ising model \cite{SeShi} is generated by
\be
\label{GR_Pafn}
U_{r-1}(Y)U_{s-1}(X) , \ \ \ \ r=1,\ldots,p-1=1,2,3,\ \ \ \ s=1,\ldots,q-1=1,2
\ee 
modulo relations
\be
\label{Is_ring}
U_{p-1}(Y) = U_3(Y) = 0
\\
U_{q-1}(X) = U_2(X) = 0 
\ee 
following from the singularities of the Chebyshev curve \eqref{Pafn_Ising}. When acting to the tachyonic module these are supplied by
\be
\label{Is_tach}
U_{p-2}(Y)=U_2(Y)=U_1(X)=U_{q-2}(X)
\ee
which actually reduces the number of independent polynomials in \eqref{GR_Pafn} to \(\frac12(p-1)(q-1)=3\).

Indeed, using the relations \eqref{Is_ring}, \eqref{Is_tach} one can reduce \eqref{GR_Pafn} to the set \(\{U_0(Y)=1, U_1(Y), U_1(X)\}\), which at the points where \(dX=4U_3(z)=0\) can be further transformed to
\be
U_1(Y) = U_1(T_3(z))  = 2T_3(z) = U_1(z)U_2(z) - 2U_1(z) = U_3(z) - U_1(z)\ \stackreb{dX=0}{=}\ -U_1(z)
\\
U_1(X) = U_1(T_4(z))  = 2T_4(z) = U_1(z)U_3(z) - 2U_2(z)   \stackreb{dX=0}{=}\ -2U_2(z)  \stackreb{dX=0}{=}\ 2U_4(z)
\ee 

\section{Residue and Verlinde formulas}

The 3-point functions in minimal gravity should reproduce the structure constants of the fusion algebra of minimal models \cite{SeShi,Zam3pt}. These structure constants are known to enter the famous Verlinde formula \cite{Ver}, relating them with the modular transformation matrices. We show in this section that Verlinde formula is actually directly related with the residue formula for the third derivatives of the Krichever tau-function.

\subsection{Residue formula on Chebyshev curves}

For the residue formula on the Chebyshev curves one gets
\be
\label{Cheres}
\F_{ijk}=\frac{pq}{ijk}\ \res_{dX=0} \frac{dH_i dH_j dH_k}{dX dY} =  \sum_{U_{p-1}(z)=0}\frac{U_{i-1}(z)U_{j-1}(z)U_{k-1}(z)}{U_{p-1}(z)U_{q-1}(z)} =
\\
=  \sum_{m=1}^{p-1}\frac{U_{i-1}(z_m)U_{j-1}(z_m)U_{k-1}(z_m)}{U_{p-1}'(z_m)U_{q-1}(z_m)} =
 \sum_{m=1}^{p-1}\frac{\sin\frac{i\pi m}p\sin\frac{j\pi m}p\sin\frac{k\pi m}p}{\sin\frac{q\pi m}p\sin^2\frac{\pi m}pU_{p-1}'(z_m)} =
\\
= \frac1p \sum_{m=1}^{p-1}(-1)^m\frac{\sin\frac{i\pi m}p\sin\frac{j\pi m}p\sin\frac{k\pi m}p}{\sin\frac{q\pi m}p}
\ee 
where we have used that \(U_{p-1}(z_m)=0\) at \(\{z_m=\cos\frac{\pi m}p|m=1,\ldots,p-1\}\) and
\be
U_n(x)'=\frac{nxU_n(x)-(n+1)U_{n-1}(x)}{x^2-1}
\ee 
i.e.
\be
U_{p-1}(z_m)'=\frac{pU_{p-2}(z_m)}{1-z_m^2} = \frac{p(-1)^m}{\sin^2\frac{\pi m}p}
\ee 
Expression in the rhs of \eqref{Cheres} appears naturally in the Verlinde formula \cite{Ver} for the minimal models (see e.g. \cite{DiFr} and references therein). Indeed, upon substitution of \(k_{r,s}= rq-sp\) into S-matrix of modular transformations for the minimal model characters
\be
S_{rs,\rho\sigma} = 2\sqrt{\frac2{pq}} (-1)^{1+s\rho+r\sigma} \sin\left(\frac{\pi q}p r\rho\right)\ \sin\left(\frac{\pi p}q s\sigma\right) = 
\\
= 2\sqrt{\frac2{pq}} 
 \sin\frac{\pi k_{r,s}\rho}p  \sin\frac{\pi k_{r,s}\sigma}q \equiv S_{\rho\sigma}(k_{r,s})
\ee 
Using this representation, the Verlinde formula can be rewritten as
\be
\mathcal{N}_{k_1k_2k_3} = \sum_{(\rho,\sigma)\in \Pi_{p,q}} \frac{S_{\rho\sigma}(k_1)S_{\rho\sigma}(k_2)S_{\rho\sigma}(k_3)}{S_{\rho\sigma}^{(0)}}
\ee 
with \(\Pi_{p,q}=\{(r,s)|r=1,\ldots,p-1;s=1,\ldots,q-1;k_{r,s}=rq-sp>0\}\) and \(S_{\rho\sigma}^{(0)}=S_{\rho\sigma}(p-q)\) corresponds to the unit operator \((r,s)\in\Pi_{p,q}:\ k_{r,s}=p-q\). Then
\be
\label{VerFactor}
\mathcal{N}_{k_1k_2k_3} =\sum_{(\rho,\sigma)\in \Pi_{p,q}} \frac{S_{\rho\sigma}(k_1)S_{\rho\sigma}(k_2)S_{\rho\sigma}(k_3)}{S_{\rho\sigma}^{(0)}} = 
\frac 12 \sum_{\rho=1}^{p-1}\sum_{\sigma=1}^{q-1} \frac{S_{\rho\sigma}(k_1)S_{\rho\sigma}(k_2)S_{\rho\sigma}(k_3)}{S_{\rho\sigma}(p-q)} =
\\
= \frac4{pq}\sum_{\rho=1}^{p-1}\sum_{\sigma=1}^{q-1}\frac{ \sin\frac{\pi k_1\rho}p  \sin\frac{\pi k_1\sigma}q\sin\frac{\pi k_2\rho}p  \sin\frac{\pi k_2\sigma}q\sin\frac{\pi k_3\rho}p  \sin\frac{\pi k_3\sigma}q}{(-1)^{1+\rho+\sigma}\sin\frac{\pi q\rho}p  \sin\frac{\pi p\sigma}q} =
\\
= -\frac4{pq}\sum_{\rho=1}^{p-1} (-1)^{\rho}\frac{ \sin\frac{\pi k_1\rho}p\sin\frac{\pi k_2\rho}p\sin\frac{\pi k_3\rho}p }{\sin\frac{\pi q\rho}p}\ 
\sum_{\sigma=1}^{q-1}(-1)^{\sigma}\frac{\sin\frac{\pi k_1\sigma}q\sin\frac{\pi k_2\sigma}q\sin\frac{\pi k_3\sigma}q}{\sin\frac{\pi p\sigma}q}
\ee 
One can therefore conclude, that
\be
\label{NFFD}
\frac14 \mathcal{N}_{ijk} = 
- \F_{ijk}\F_{ijk}^D
\ee 
for any admissible triples \(\{ijk\}\in \Pi_{p,q}^{\otimes 3}\), with \(\F_{ijk}\) defined in \eqref{Cheres} and \(\F_{ijk}^D\) being the same expression for the 
formally ``pq-dual'' 3-point function~\footnote{Here we discuss this expression only at the ``conformal background'' or on the Chebyshev curves, its exact relation with the dual Krichever tau-function is a non-trivial issue.}
\be
\label{CheresD}
\F_{ijk}^D = \frac{pq}{ijk}\ \res_{dY=0} \frac{dH_i dH_j dH_k}{dX dY}
\ee 

\subsection{Proof in \((p,q)=(2K+1,2)\) case}

For particular series \((p,q)=(2K+1,2)\) the r.h.s. of \eqref{VerFactor} simplifies, since the second factor at \(q=2\) (and therefore \(\sigma=1\), while all \(\{k_i=2l_i+1\}\) are odd) just gives
\be 
\sum_{\sigma=1}^{q-1}(-1)^{\sigma}\frac{\sin\frac{\pi k_1\sigma}q\sin\frac{\pi k_2\sigma}q\sin\frac{\pi k_3\sigma}q}{\sin\frac{\pi p\sigma}q}\ \stackreb{ \{k_i=2l_i+1\}}{=}\ 
- \frac{\sin\pi \left(l_1+\frac12\right)\sin\pi \left(l_2+\frac12\right)\sin\pi \left(l_3+\frac12\right)}{\sin\pi \left(K+\frac12\right)} =
\\
= (-1)^{1+K+l_1+l_2+l_3}
\ee 
Formula \eqref{NFFD} now reads
\be
 \mathcal{N}_{ijk} = 2(-1)^{K+l_i+l_j+l_k} \F_{ijk}
\ee 
and it is convenient now to parameterize \(k_i\equiv k_{{r_i},1}=2r_i-2K-1\ \stackreb{r_i=K+l_i+1}{=}\ 2l_i+1\), \(l_i=0,\ldots,K-1\) and \(t_{\tilde{i}}=t_{2l_i+1}=t_{2(K -\tilde{i})+1}\) with
\(\tilde{i}=1,\ldots,K\). Then we get from \eqref{Cheres}
\be
\F_{\tilde{i}\tilde{j}\tilde{k}} = \frac1{2K+1} \sum_{m=1}^{2K}(-1)^m\frac{\sin\frac{k_i\pi m}{2K+1}\sin\frac{k_j\pi m}{2K+1}\sin\frac{k_k\pi m}{2K+1}}{\sin\frac{2\pi m}{2K+1}} =
\\
=  \frac2{2K+1} \sum_{m=1}^{K}(-1)^m\frac{\sin\frac{k_i\pi m}{2K+1}\sin\frac{k_j\pi m}{2K+1}\sin\frac{k_k\pi m}{2K+1}}{\sin\frac{2\pi m}{2K+1}} =
\\
= - \frac2{2K+1} \sum_{m=1}^{K}\frac{\sin\frac{2\pi m\tilde{i}}{2K+1}\sin\frac{2\pi m\tilde{j}}{2K+1}\sin\frac{2\pi m\tilde{k}}{2K+1}}{\sin\frac{2\pi m}{2K+1}}
\ee 
and
\be
\label{KriVerp2}
\mathcal{N}_{\tilde{i}\tilde{j}\tilde{k}} = 2(-1)^{\tilde{i}+\tilde{j}+\tilde{k}} \F_{\tilde{i}\tilde{j}\tilde{k}} = (-1)^{1+\tilde{i}+\tilde{j}+\tilde{k}}\frac4{2K+1} \sum_{m=1}^{K}\frac{\sin\frac{2\pi m\tilde{i}}{2K+1}\sin\frac{2\pi m\tilde{j}}{2K+1}\sin\frac{2\pi m\tilde{k}}{2K+1}}{\sin\frac{2\pi m}{2K+1}}
\ee 
In particular, it is easy to see from this formula that \(\mathcal{N}_{1\tilde{i}\tilde{j}} =\delta_{\tilde{i}\tilde{j}} \) while \(\F_{1\tilde{i}\tilde{j}}\) is unit matrix up to a normalization constant.

Actually, for this series, the Krichever residue formula gives a proof of Verlinde's formula, since the residue \eqref{Cheres} can be also computed by moving the contour to other singularities at \(dY=0\) or \(z=0\) and \(z=\infty\). One finds then for \eqref{KriVerp2}, that
\be
\label{Verproof2}
(-1)^{1+\tilde{i}+\tilde{j}+\tilde{k}}\mathcal{N}_{\tilde{i}\tilde{j}\tilde{k}} = \frac4{2K+1} \sum_{m=1}^{K}\frac{\sin\frac{2\pi m\tilde{i}}{2K+1}\sin\frac{2\pi m\tilde{j}}{2K+1}\sin\frac{2\pi m\tilde{k}}{2K+1}}{\sin\frac{2\pi m}{2K+1}} = 
\\
= - \res_{U_{2K }(z)=0}\frac{U_{2l_i}(z)U_{2l_j}(z)U_{2l_k}(z)dz}{ z U_{2K }(z)} =
\left(\res_{z=0}+\res_{z=\infty}\right)\frac{U_{2l_i}(z)U_{2l_j}(z)U_{2l_k}(z)dz}{ z U_{2K }(z)} =
\\
=\left(\res_{w=i}+\res_{w=0}\right)\frac{\left(w^{2{l_i}+1}-w^{-2{l_i}-1}\right) \left(w^{2{l_j}+1}-w^{-2{l_j}-1}\right)
	\left(w^{2{l_k}+1}-w^{-2{l_k}-1}\right)}{\left(w^{2}-w^{-2}\right)\left(w^{2K +1}-w^{-2K -1}\right)}\frac{dw}{w}
\ee 
after the change of variable \(z=\frac12\left(w+\frac1w\right)\). Hence
\be
\label{Verproof2}
\mathcal{N}_{\tilde{i}\tilde{j}\tilde{k}} =
\\
= 1 + (-1)^{1+\tilde{i}+\tilde{j}+\tilde{k}}\res_{w=0}\frac{\left(w^{2{l_i}+1}-w^{-2{l_i}-1}\right) \left(w^{2{l_j}+1}-w^{-2{l_j}-1}\right)
	\left(w^{2{l_k}+1}-w^{-2{l_k}-1}\right)}{\left(w^{2}-w^{-2}\right)\left(w^{2K +1}-w^{-2K -1}\right)}\frac{dw}{w}\ \stackreb{t=w^2}{=}\ 
\\
= 1 + (-1)^{1+\tilde{i}+\tilde{j}+\tilde{k}}\res_{t=0}t^{\tilde{i}+\tilde{j}+\tilde{k}-2K }\sum_{l=0}^{K -2}t^{2l}\left(1-t^{2 (K-\tilde{i}) + 1}-t^{2 (K-\tilde{j}) + 1}-t^{2 (K-\tilde{k}) + 1}\right)\frac{dt}{t} =
\\
= 1 - \sum_{l=0}^{K -2}\delta_{\tilde{i}+\tilde{j}+\tilde{k}+2l,2K } -  \sum_{l=0}^{\left[\frac{K -3}{2}\right]}\delta_{\tilde{i}+\tilde{j}+2l+1,\tilde{k}} -  \sum_{l=0}^{\left[\frac{K -3}{2}\right]}\delta_{\tilde{i}+\tilde{k}+2l+1,\tilde{j}} -  \sum_{l=0}^{\left[\frac{K -3}{2}\right]}\delta_{\tilde{j}+\tilde{k}+2l+1,\tilde{i}} 
\ee
The last expression is not very convenient for practical purposes, but it can be rewritten in the form
\be
\label{Verproof3}
\mathcal{N}_{\tilde{i}\tilde{j}\tilde{k}} =   \sum_{l=0}^{\min(\tilde{i},\tilde{j})-1}\delta_{|\tilde{i}-\tilde{j}|+2l+1,\tilde{k}} +  \sum_{l=0}^{\left[\frac{K -2}{2}\right]}\delta_{\tilde{i}+\tilde{j}+\tilde{k},2(K +l+1)} 
\ee

\subsection{Non-algebraic generalization}	

It is interesting to point out, that similar reasoning can be applied to	``continuous'' theory (see e.g. \cite{CEMR}) where the analog of the formulas \eqref{VerFactor} and, especially \eqref{KriVerp2} looks as~\footnote{Actually this formula corresponds to the expansion of Zamolodchikov's expression from \cite{Zam3pt}.}
\be 
\label{cont3pt}
\mathcal{N}(p_1,p_2,p_3) = 2b\sum_{m=1}^\infty(-1)^m\frac{	\sin 2\pi m bp_1 \sin 2\pi m bp_2 \sin 2\pi m bp_3}
{\sin \pi m b^2}  
\ee 
depending now on three continuous variables \(\{p_{1,2,3}\}\). Here
	\be 
b^2 \in i\mathbb{R}
\ee 
instead of \(b^2=\frac{q}{p}\) (we have actually \(0<b^2<1\)) in minimal theory. Note also, that
	\be 
	(-1)^m\sin \pi m b^2  = \sin (\pi m b^2 + \pi m) = \sin 2\pi m b p_0 = \mathcal{S}_0^m
	\ee 
where the Liouville momentum \(p_0 = (1/b+b)/2\) corresponds to \(h_0=0\) or ``unity'' operator. 
	
In our context it is important to understand, that formula \eqref{cont3pt} actually
	comes from residue formula \eqref{residue} for a non-algebraic curve (cf. with \eqref{Chebuni})
	\be
	X(\mathsf{\mathsf{z}}) = \cos  \pi b^{-1} \mathsf{z},\ \ \ \ Y(\mathsf{z}) = \cos\pi b \mathsf{z}
	\ee 
which obeys trigonometric uniformization, but no similar to \eqref{Chebcu} polynomial representation. Indeed	
	\be 	
	\mathcal{N}(p_1,p_2,p_3) = \sum_{dX=0} \frac{dH_{p_1}dH_{p_2}dH_{p_3}}{dX\ dY} = \sum_{X'(\mathsf{z})=0} \frac{\phi(p_1 \mathsf{z})\phi(p_2 \mathsf{z})\phi(p_3 \mathsf{z})}{X''(\mathsf{z})Y'(\mathsf{z})}
	\ee
	since \(X'(\mathsf{z})\sim \sin  \pi b^{-1} \mathsf{z} =0\) at \(\mathsf{z}_m= b m\), \(m\in \mathbb{Z}\), where
	\be
	Y'(\mathsf{z}_m) \sim  b\sin\pi b \mathsf{z}_m = b\sin \pi b^2 m,\ \ \ \ \ X''(\mathsf{z}_m)\sim b^{-2}\cos  \pi b^{-1} \mathsf{z}_m =  b^{-2}(-1)^m
	\ee 
	and the rest comes from identification \(\phi (p\mathsf{z}) = \sin 2\pi p\mathsf{z}\), just as in the case of Chebyshev \(U\)-polynomials \eqref{fimm}.

\section{Discussion}

We have discussed here a particular application of the quasiclassical Krichever tau-function to the minimal gravity in conformal background, and shown that the universal residue formula \cite{KriW} for its third derivatives or 3-point functions in the conformal background with only \(\mu\neq 0\) turns to be directly related with the Verlinde formula for minimal models of 2d CFT. For the simplest \((p,q)=(2K+1,2)\) series the residue formula gives actually a direct analytic proof of the Verlinde formula \cite{Ver} and established an exact equivalence between the Liouville world sheet and ``integrable'' method of the computation of the (very restricted class of) correlation numbers in minimal 2d gravity.

Unfortunately the approach of \cite{KriW} does not provide us with simple explicit formulas for more sophisticated correlators, starting already from the multipoint functions on sphere (even in conformal background), which do not have as simple structure and dependence upon quantum numbers, as the 3-point functions. The relation with Verlinde formula allows only to get similar to \eqref{KriVerp2}, \eqref{cont3pt} expressions for some of their building ingredients, being the ``conformal block's numbers'' \(\mathcal{N}_{i_1,\ldots,i_n}\) in corresponding minimal model. In particular, the symmetricity of \(\mathcal{N}_{i_1,\ldots,i_4}\) can be seen in this context as a consequence of WDVV equations.

The case of generic \((p,q)\)-minimal string deserves further investigation even at this very restricted level. It turns out, however, that factorization of the Verlinde formula into two dual 3-point functions \eqref{NFFD} expressed through the residue formulas, may give an extra hint towards many recent attempts to find an axiomatic ``target-space'' formulation of computation these correlation numbers, hopefully to be more understandable than topological recursion. In particular, it is nice to point out, that this ``residue-Verlinde'' parallels already show up in certain models of non-minimal 2d gravity. Since generally Krichever tau-function appears in quasiclassical formulation of far more wide class of (even multidimensional!) theories, it could be indeed a sign of possible existence of some generic axiomatic formulation.

\noindent
{\bf Acknowledgments}.  I am deeply indebted to A.~Artemev for permanent illuminating discussions, sharing his insights and many useful comments, and to P.~Gavrylenko for important technical suggestions and remarks. I would like to thank the organizers of the \emph{XIV Workshop on Geometric Correspondences of Gauge Theories} at SISSA and the School-Conference \emph{Integrable Systems and Algebraic Geometry} at BIMSA, where the preliminary results were first presented.

\end{document}